# Space-Efficient Simulation of Quantum Computers

Michael P. Frank
Dept. of Electrical & Computer Eng.
FAMU-FSU College of Engineering
2525 Pottsdamer St., Rm. A341
Tallahassee, FL  32310, USA
+1 (850) 410-6455
mpf@eng.fsu.edu

Uwe H. Meyer-Baese
Dept. of Electrical & Computer Eng.
FAMU-FSU College of Engineering
2525 Pottsdamer St., Rm. A354
Tallahassee, FL  32310, USA
+1 (850) 410-6220
umb@eng.fsu.edu

Irinel Chiorescu
National High Magnetic Field Lab
Florida State University
1800 East Paul Dirac Drive
Tallahassee, FL  32310, USA
+1 (850) 644-1726
ic@magnet.fsu.edu

Liviu Oniciuc
Dept. of Electrical & Computer Eng.
FAMU-FSU College of Engineering
2525 Pottsdamer St., Rm. A341
Tallahassee, FL  32310, USA
+1 (850) 410-6455
oniciuc@eng.fsu.edu

Robert A. van Engelen
Dept. of Computer Science
Florida State University
160 James J. Love Bldg.
Tallahassee, FL  32306, USA
+1 (850) 645-0309
engelen@cs.fsu.edu

## ABSTRACT
Traditional algorithms for simulating quantum computers on classical ones require an exponentially large amount of memory, and so typically cannot simulate general quantum circuits with more than about 30 or so qubits on a typical PC-scale platform with only a few gigabytes of main memory.  However, more memory-efficient simulations are possible, requiring only polynomial or even linear space in the size of the quantum circuit being simulated.  In this paper, we describe one such technique, which was recently implemented at FSU in the form of a C++ program called SEQCSIM, which we releasing publicly.  We also discuss the potential benefits of this simulation in quantum computing research and education, and outline some possible directions for further progress.

## Categories and Subject Descriptors
F.1.2 [**Modes of Computation**]: Probabilistic computation.

## 1. INTRODUCTION
Quantum computing [1] is a fundamentally new abstract model of computation developed in the 1980s and '90s for exploring the theoretical capabilities of hypothetical "quantum computers" (not yet implemented at useful scales) which would exhibit and exploit exotic quantum-mechanical phenomena (such as superposition of states, interference effects and entanglement) in their fundamental logical mode of operation.  The study of quantum computers is of significant intrinsic academic interest because, in principle, quantum computers would provide exponential speedups on several important classes of problems, including factoring of large numbers [2] (useful in cryptanalysis) and simulating the quantum-mechanical behavior of physical systems [3] (*e.g.*, atoms, molecules, and nanoscale devices).

Occasionally in the quantum computing literature (and often in press reports), one sees claims that to simulate a quantum computer on a classical one requires an amount of memory that increases exponentially with the size of the quantum circuit being simulated.  Although this is true if the quantum state vector is represented explicitly (*e.g.* in an array), simulating the measurable statistical behavior of a quantum circuit does not actually require such an explicit representation, and so much more space-efficient simulations are possible.

A simple, general algorithmic transformation for trading off space for time which has long been known in computational complexity theory (and which can be applied to any computation with a polynomial-depth dataflow graph) implies that any polynomial-time quantum algorithm requires only polynomial space to simulate classically, yielding the basic complexity theoretic relation that BQP (the class of problems solvable in probabilistic polynomial time on a quantum computer) is a subset of PSPACE [4].  However, there are few (if any) publicly-available quantum computer simulators leveraging this important insight.

To help remedy this situation, at Florida State University we recently developed (in ANSI C++) a working prototype of a new quantum computer simulator called SEQCSIM (say "SEEK-sim"), standing for <u>S</u>pace-<u>E</u>fficient <u>Q</u>uantum <u>C</u>omputer <u>S</u>imulator. SEQCSim uses only *linear* space in the size of the quantum circuit being simulated.  More precisely, for $s$ bit wide, $t$ gate circuits, SEQCSIM's space usage grows only as $O(s + t)$.

If desired, this can be reduced further to only $O(s + k)$, where $k$ is the maximum number of nontrivial operations (for a certain definition of "nontrivial") in any qubit's predecessor graph.  For many of the important quantum algorithms that have been described (such as Shor's algorithm [2]), $k$ itself is only $O(s)$, and so in these cases, the overall space complexity reduces to just $O(s)$, *i.e.*, only proportional to the space usage of the quantum computer being simulated. In other words, available memory need not significantly limit the size of the quantum circuits we can simulate.

Execution time is still a limiting factor (still exponential, in the worst case), but to the extent that additional memory may be readily available, it can be used in a straightforward way to boost the performance of our simulator. In addition, our simulator can be implemented on a single, very fast, FPGA chip, requiring no slow off-chip accesses to external memory, and using special-purpose parallel hardware to further boost speed. (We are presently working on a custom architecture of this sort, which is expected to improve the performance of our simulator by ~50×.)

In addition to its advantages in terms of computational complexity, our simulation technique also provides an interesting pedagogical illustration of several important conceptual aspects of quantum mechanics.

For example, David Bohm's interpretation [5] of quantum mechanics shows that, contrary to widespread belief, we can treat quantum systems as always possessing a definite classical state, which tracks the (local) flow of probability mass through the system's configuration space. This picture contrasts with the inherently indefinite state pictured in other interpretations. Our simulation leverages Bohm's insight by following, in each run, only a single classical (computational basis) state, which is evolved consistently with the flow of probability mass implied by the quantum algorithm. This saves memory and computational effort that might otherwise be spent considering final states that would end up with zero (or negligible) aggregate probability.

Another reformulation of quantum mechanics illustrated in our simulator is that of Feynman [6], who showed that the amplitudes of quantum wavefunctions can be calculated using a *path integral*, basically a sum ranging over the possible trajectories through configuration space that may be taken by the system. The advantage of this approach is that trajectories can be explored sequentially, so that the entire wavefunction never needs to be explicitly represented. This is another way of explaining what allows our simulation to run in linear space.

Finally, via the particular way in which it combines Bohm's and Feynman's pictures, SEQCSIM gives us a new way of conceptualizing the universe described by quantum mechanics, as one in which each system has just a single classical state that evolves under the influence of not only the present local variables, but also a hidden memory of the structure of past local interactions (quantum logic gates) in the causally-connected history of the system. This picture provides a compact hidden-variables model of quantum mechanics which partially addresses old philosophical objections to quantum mechanics raised by Einstein and others [7]. (The hidden information is non-local in the sense required to avoid Bell's "no-hidden-variables" theorem [8].)

We are currently developing a new version of our simulator illustrating this new conceptual picture even more explicitly, by providing a novel C++ API that lets programmers construct and manipulate class objects that act exactly like real-world qubits (just more slowly). By looking "behind the scenes" at the simulator code, scholars can see for themselves how, with help of some computational effort, all of the supposedly-weird behavior of quantum systems can arise from a classical model, given a modest-sized record of past local interactions.

To preview the rest of this paper: Section 2 outlines and explains the current simulator algorithm. Section 3 describes a simple example circuit. Section 4 briefly describes what we have in mind for our forthcoming C++ API and FPGA-based hardware accelerator. Section 5 concludes.

**Listing 1. Outline of the algorithm used in the present SEQCSim quantum computer simulator.** (Some details omitted.)

```
procedure SEQCSim::run():
    curState := inputState;       // Current basis state
    curAmp := 1;                  // Current amplitude
    for PC =: 0 to #gates,        // Current gate index
        (w.r.t. gate[PC] operator and its operands,)
        for each neighbor nbr_i of curState,
            if nbr_i = curState, amp[nbr_i] := curAmp;
            else amp[nbr_i] := calcAmp(nbr_i);
        amp[] := opMatrix * amp[]; // Matrix prod.
        // Calculate probabilities as normalized
        // squares of amplitudes.
        prob[] := normSqr(amp[]);
        // Pick a successor of the current state.
        i := pickFromDist(prob[]);
        curState := nbr_i;  curAmp := amp[nbr_i].

// Recursive amplitude-calculation procedure
function SEQCSim::calcAmp(Neighbor nbr):
    curState := nbr;
    if PC=0 return (curState = inputState) ? 1 : 0;
    (w.r.t. gate[PC−1] operator and its operands,)
    for each predecessor pred_i of curState,
        PC := PC − 1;
        amp[pred_i] = calcAmp(pred_i);
        PC := PC + 1;
    amp[] := opMatrix * amp[];
    return amp[curState];
```

## 2. SEQCSIM ALGORITHM

The current version of our simulator (v0.8) uses a very simple algorithm. Given input files specifying a quantum circuit and a classical input state (*i.e.* a basis state in the computational basis), SEQCSim simply applies the quantum gates (unitary operations) one at a time, calculating at each step the amplitudes and probabilities of the possible gate outputs, and selecting an "actual" output state at random, according to the amplitude distribution, using a standard PSRG (pseudo-random number generator, which should be re-seeded on each run if multiple runs are needed). The only difficult part occurs when the quantum gate is *nontrivial* for the given input state, meaning that its unitary matrix has $b>1$ entries in the selected column. In this case, the output amplitudes depend not only on the amplitude of the actual input state, but also on that of $b-1$ small Hamming-distance neighbors of the input state, varying from it at bits that are operands of the current gate.

Calculating the neighbor-state amplitudes involves recursively calculating the amplitudes of the neighbors' immediate predecessors (the consistent input states to the *previous* gate operation) in the same way. The recursion bottoms out at the start of the quantum circuit, where the state identical to the input state is assigned amplitude 1 and all other states are assigned amplitude 0. Listing 1 above gives brief pseudocode for the algorithm.

The space complexity analysis is as follows. Given gates of small constant arity (# of qubits) $a$, specifying a neighbor state to visit only requires $a = \Theta(1)$ bits; these "delta bit vectors" to near-by states (and a constant amount of other stack frame data) are all that actually needs to be pushed onto the stack at each step of the recursion; at full depth, the stack contents give a trajectory through configuration space from time 0 to the current top-level PC value; this has size $\Theta(t)$, if $t$ is the number of gates. (The recursive procedure traverses these trajectories successively and effectively computes the path integral.) Meanwhile, *curState* has size $\Theta(s)$, if $s$ is the width of the quantum circuit, and all other variables and arrays are only constant size (approximating logarithmic factors as constant), and so total space usage is $\Theta(s + t)$.

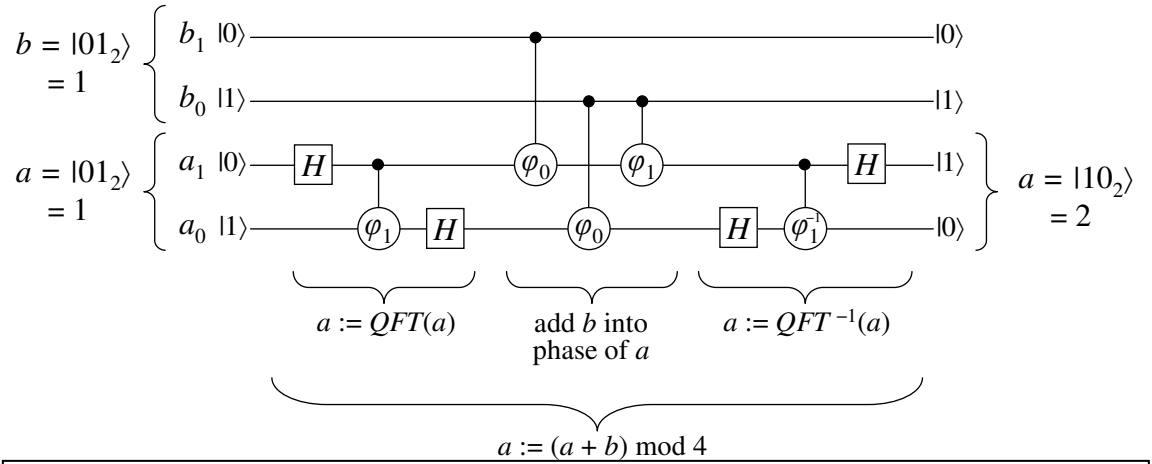

**Figure 1. A simple example quantum circuit.** Uses the quantum Fourier transform (QFT) and its inverse QFT$^{-1}$ to add two 2-bit input integers in a temporary phase space representation. Here it is computing 1 + 1 = 2.

The time complexity analysis is similarly easy. The run time is dominated by the calcAmp() recursion for the last nontrivial gate. The branching factor $b$ at each node of the recursive call tree is given by the number of predecessor states of the current state, which is equal to the rank of the block (submatrix) of the current gate operator corresponding to that state. We have $b = 1$ in the case of "trivial" gates, defined as gates whose unitary matrices can be diagonalized in the computational basis, such as classical gates and phase gates, and we have $b = 2^a$ in the worst case, for general (nontrivial) unitary gates of arity $a$. Many gates have intermediate branching factors, such as the "controlled-controlled-… rotation" gate C$^n U$, for arbitrary $n$ and 1-bit unitary $U$, which has $b=1$ in $2^{n+1}-2$ of the input cases, and $b=2$ in only the remaining 2 input cases, when all $n$ of the control bits are simultaneously 1 (and $U$ is not diagonal).

For simplicity, let $b = \Theta(1)$ be the maximum branching factor (maximum block rank) over all gates in the quantum circuit, and let $k$ be the number of nontrivial gates in the circuit; then the number of leaf nodes of the recursion is $O(b^k) = O(2^{O(k)})$. In the worst case, the deepest part of the recursion may involve traversing $t - k = O(t)$ trivial (*e.g.* classical) gates, and so the worst-case time complexity of the recursion is $\Theta(t \cdot 2^{O(k)})$. Constructing the $s$-bit initial state takes time $\Theta(s)$, but subsequent state manipulations and comparisons can be done incrementally, and need only take amortized constant time, and so total worst-case time complexity for a well-optimized implementation can be as low as $\Theta(s + t \cdot 2^{O(k)})$.

We should note that, if the number $k$ of nontrivial gates is larger than the number of qubits $s$, this approach can in fact be slower than the conventional approach of simply tracking the full quantum state vector of $2^s$ elements, which takes time $\Theta(t \cdot 2^s)$. The conventional approach can be thought of as a dynamic-pro-gramming performance optimization of our approach, caching all state amplitudes so as to avoid the work of dynamically recalculating their values whenever needed.

However, in cases where the cost for the $\Theta(2^s)$ size memory required for a traditional state-vector approach would be prohibitive, our approach provides a space-time tradeoff that can permit the simulation of quantum circuits that would be too large to simulate with conventional approaches. Furthermore, our simulator can be easily modified to perform a conventional simulation (with a sparse state vector representation) until the limit of the available memory is reached, and revert to the path-integral approach only for subsequent steps in the quantum circuit beyond that point (*i.e.*, using the last explicitly representable state vector as the initial state for the remaining part of the circuit). In this way, the simulator can take full advantage of the available memory to improve its performance on nontrivial circuits, while not limiting the size of such circuits that it can handle.

## 3. EXAMPLE QUANTUM CIRCUIT

To illustrate the operation of our algorithm, Figure 1 gives a simple example of a quantum circuit, using the standard graphical notation of quantum logic networks [1]. The strings in |⟩ brackets label classical basis states, and the icons represent quantum gates (unitary operations). $H$ is a 1-bit gate called the Hadamard transform; in terms of Pauli spin operators, it can be written as $(\sigma_x + \sigma_z) \cdot 2^{-1/2}$ = [1, 1; 1, −1]/$\sqrt{2}$, a 1-line notation for the matrix

$$H = \frac{1}{\sqrt{2}} \begin{bmatrix} 1 & 1 \\ 1 & -1 \end{bmatrix}. \quad (1)$$

Meanwhile, $\varphi_q$ is the "controlled-phase" gate for a relative-phase rotation (between |0⟩ and |1⟩ states) of $1/2^q$ of a half-circle; this is a trivial gate that can be written algebraically using rank-2 operators of identity $\hat{I}$ = [1, 0; 0 1], number $\hat{n} = \hat{a}^\dagger \hat{a}$ = [0, 0; 0, 1] (with $\hat{a}$ = [0, 1; 0, 0] the annihilation operator, $\hat{a}^\dagger$ its adjoint), number complement ($\bar{n} = \hat{a}\hat{a}^\dagger = \hat{I} - \hat{n}$), and tensor product $\otimes$,

$$\varphi_q = \bar{n} \otimes \hat{I} + \hat{n} \otimes \exp(i\pi\hat{n}2^{-q}), \quad (2)$$

or more explicitly in matrix form, as follows:

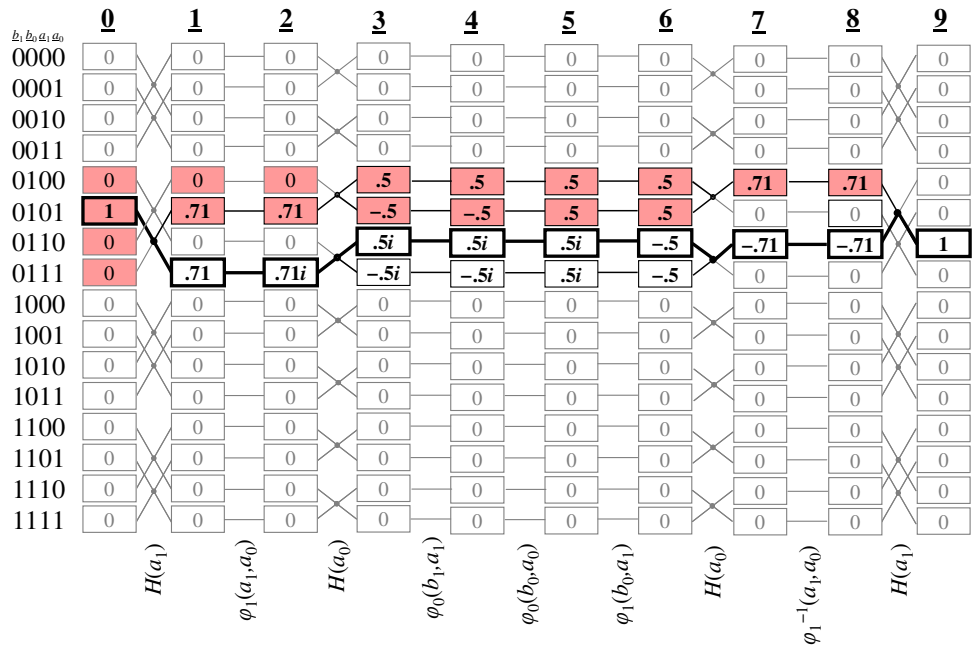

**Figure 2. State graph of figure 1 circuit.** Rows correspond to basis states; columns are the full state vectors. Lines between states show paths along which amplitude can flows as a result of each gate. (Note that the trivial φ gates produce no branching.) SEQCSim follows a single randomly-selected trajectory; the bold path is one possibility. To calculate transition probabilities for nontrivial gates, the amplitudes of neighbor states are computed using a path integral. The states shaded red are traversed by the space-efficient recursive calculation of the neighbor state at step 8, needed to calculate the output probabilities for the last *H* operation.

$$\varphi_q = \begin{bmatrix} 1 & 0 & 0 & 0 \\ 0 & 1 & 0 & 0 \\ 0 & 0 & 1 & 0 \\ 0 & 0 & 0 & \exp(i\pi 2^{-q}) \end{bmatrix} \quad (3)$$

The circuit shown uses an algorithm by Draper [9] to add a pair of 2-bit numbers *a*, *b* in-place (modulo 4) in the *a* register without using any carry bits, by first transforming *a* to a phase representation using a 2-bit quantum version of the discrete Fourier transform (QFT), adding *b* into the relative phases of *a* using the $\varphi_q$ gates in the middle of the circuit, and transforming the result in *a* back to the usual binary encoding using the inverse of the QFT circuit. ($H = H^{-1}$ is its own inverse.)

Figure 2 illustrates the operation of SEQCSim for this example. The full state vectors (columns of figure 2) are never generated. Instead, a random trajectory (say the one lined in bold) is traversed, consistently with state probabilities. To capture interference effects, at each step where a nontrivial operation takes place, appropriate neighbor states are examined, and their amplitudes calculated recursively. The states visited during this calculation for state 0110's single neighbor at step 8 are shaded red.

Figure 3 below shows the actual input files (trimmed somewhat for brevity) used to code the above example in SEQCSim's current file format, and the actual output produced currently by the simulator. The current input and output formats are just ad-hoc temporary solutions, designed primarily for our internal use in testing & debugging of the simulator.

## 4. FUTURE WORK

The present simulator is a bit cumbersome to use since the quantum circuit to be simulated must be specified explicitly in the low-level text input format of figure 3. Such descriptions can become rather tedious. For ease of constructing larger circuits, we would prefer a quantum programming language for more abstractly describing complex quantum algorithms, like in a normal programming language. A number of quantum programming languages have been described previously (see [10] for a survey), but few of them are both easy to use and readily accessible.

One particularly simple approach we are exploring (inspired by [11]) is to let the quantum programming language be a conventional OOP (object-oriented programming) language such as C++, with a class library API that emulates the behavior of real qubits, invoking the simulator "behind the scenes," as it were.

The implementation of this approach leads to some interesting considerations. The quantum circuit can be constructed dynamically and incrementally by the API as the programmer creates qubit objects and applies quantum operations to them. Rather than treating quantum states as monolithic non-local objects, the simulator can naturally confine its attention to the qubits currently being manipulated, and the causally-connected graph of gates in their history. This approach facilitates optimizations that permit us to ignore gates that are not causally connected to current qubits, and to prune recursive trajectories as soon as they are found inconsistent with earlier partial measurements or with the initial state (rather than waiting until no more gates can be undone). A future paper will fully describe the new algorithm.

Later, the simulation environment can be extended to do even more aggressive optimizations, such as applying algebraic transformations to dynamically restructure the past computation graph to a form permitting faster simulation. This approach may be thought of as an optimizing JIT (just-in-time) compilation of the quantum algorithm.

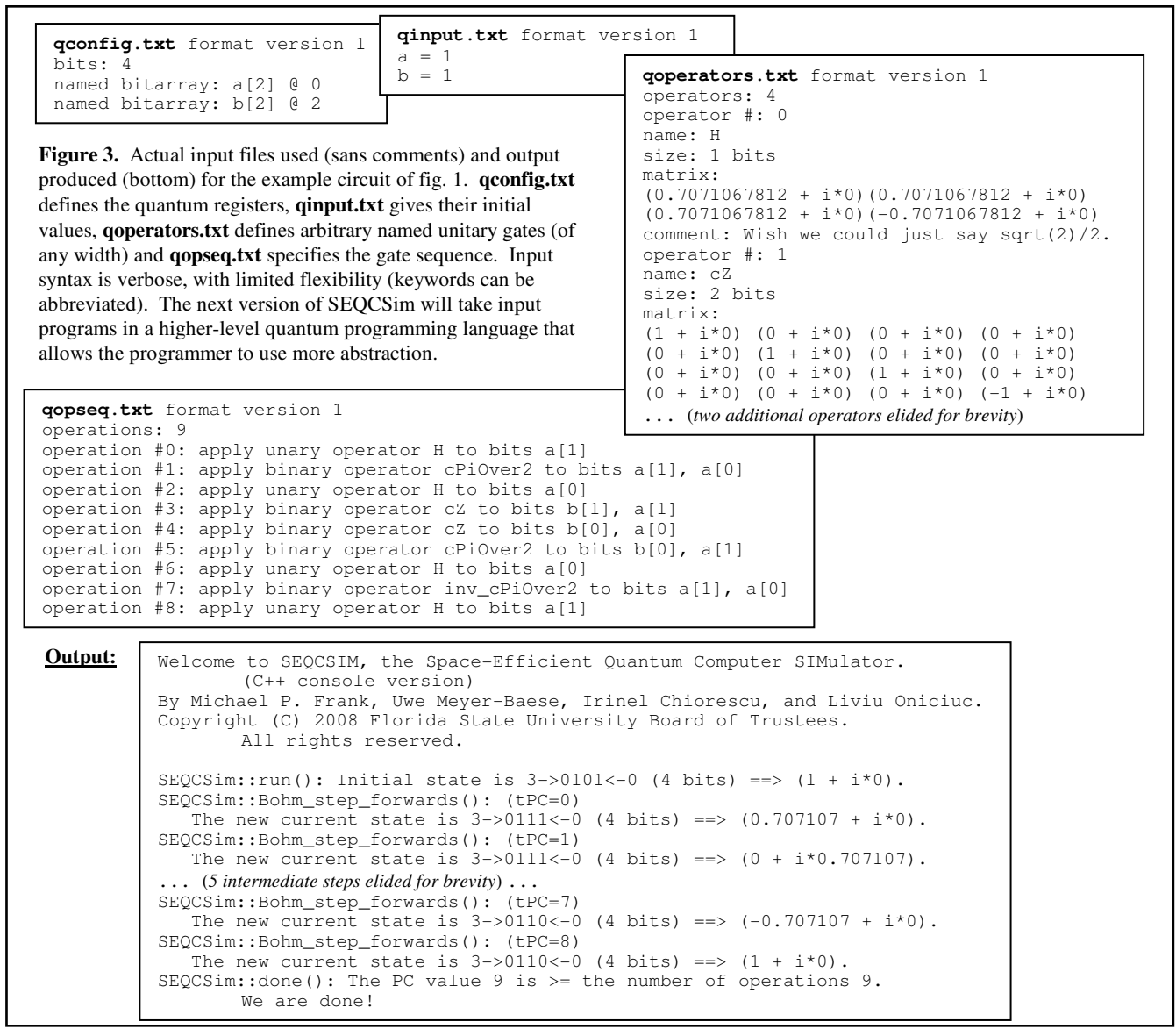

**Figure 3.** Actual input files used (sans comments) and output produced (bottom) for the example circuit of fig. 1. **qconfig.txt** defines the quantum registers, **qinput.txt** gives their initial values, **qoperators.txt** defines arbitrary named unitary gates (of any width) and **qopseq.txt** specifies the gate sequence. Input syntax is verbose, with limited flexibility (keywords can be abbreviated). The next version of SEQCSim will take input programs in a higher-level quantum programming language that allows the programmer to use more abstraction.

Finally, once the design of the simulator algorithm itself has stabilized, we also intend to implement, in an FPGA platform, a special-purpose hardware architecture to significantly speed up the simulation by a constant factor of ~50-100×. This will be done by using custom high-bandwidth memory structures and parallel arithmetic datapaths to reimplement in hardware the kernel of the time-consuming recursive part of the simulation algorithm. The higher-level control can remain in software, running on an embedded RISC soft-core (such as Altera's NIOS or Xilinx's MicroBlaze), augmented with new special-purpose instructions to invoke the custom hardware. Further, numerous branches of the recursive path-integral computation (or the top-level stochastic simulation) can be performed in parallel on a single large FPGA chip, boosting performance even further.

## 5. CONCLUSION

We have implemented and verified a working quantum computer simulator that requires only an amount of memory that grows in linear proportion to the size of the quantum circuit being simulated. The simulator still requires worst-case exponential time, but its performance can be improved in ways we are pursuing.

The structure of our simulation algorithm (and its future version in development) suggests an interesting new ontological interpretation of quantum mechanics, which may help to dispel the philosophical unease sometimes associated with traditional interpretations – since the simulation requires no "magic," nor exponentially large numbers of parallel universes; rather, just qubits with classical states that also carry with them a link into the graph of interactions in their causally-connected history.

SEQCSim (especially the future object-library version) is thus potentially useful as a tool for teaching scholars about quantum computing and the broad computational picture of quantum mechanics that it offers. One can construct quantum circuits that model "weird" quantum phenomena such as non-local entanglement, "teleportation," *etc.*, and understand how these circuits work in terms of an underlying simulation that is entirely classical and locally generated (if rather time-consuming).

If we can accept a mental picture of the universe as working analogously to our simulator, *i.e.*, doing a complex graph computation behind the scenes each time it updates a particle's state, then we need not find nature's quantum-mechanical behavior to be particularly mysterious or paradoxical any longer. So perhaps Feynman's famous lament that "no one understands quantum mechanics" [12] can finally be put to rest.

## 6. ACKNOWLEDGMENTS
This work was graciously supported by a planning grant from the Council on Research and Creativity of the Office of Research at Florida State University.